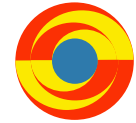

# Self-Consistent Study of the Superconducting Gap in the Strontium-doped Lanthanum Cuprate

**Pedro Contreras**[1, *], **Dianela Osorio**[2], **Anjna Devi**[3]

[1]Department of Physics, University of Los Andes, Mérida, Venezuela

[2]Department of Brain and Behavioral Sciences, University of Pavia, Pavia, Italy

[3]Department of Physics, Himachal Pradesh University, Shimla, India

**Email address:**

pcontreras@ula.ve (Pedro Contreras), daobfisica@gmail.com (Dianela Osorio), anjnahpu90@gmail.com (Anjna Devi)

[*]Corresponding author

**Abstract:** This work is aimed at numerically investigating the behavior of the Fermi energy in Strontium-doped Lanthanum Cuprate, using a numerical zero temperature elastic scattering cross-section procedure in the unitary collision regime. The main task is to vary the zero temperature superconducting energy gap from its zero value in the normal state, to the highest value of 60 meV. We find that there are two different reduced phase space regimes for the first harmonic line node's order parameter. The first one, represented by a Fermi energy with a value of - 0.4 meV, where the rest of the material parameters, and the degrees of freedom of the normal to the superconducting phase transition are not sensitive to the self-consistent variation of the zero temperature superconducting energy gap. A different case is found when in the self-consistent numerical procedure, the Fermi energy takes a value of - 0.04 meV, indicating that the fermion-dressed quasiparticles have material parameters strongly sensitive to the numerical changes in the zero temperature gap, resulting in a reduced phase space, where the input and output zero superconducting energy values, and the degrees of freedom are separated by the self-consistent numerical analysis. The first scenario considers that when the Fermi energy and the nearest hopping terms have the same order of magnitude, the physics can be described by a picture given by nonequilibrium statistical mechanics. A second scenario indicates, that when the Fermi energy parameter and the hopping term have different order of magnitude; the physical picture tends to be related to the nonrelativistic quantum mechanical degrees of freedom coming from quasi-stationary quantum energy levels, with a damping term seen in the probability density distribution function, that is described in the configuration space. Henceforth, it is concluded that the use of the zero temperature elastic scattering cross-section links the phase and configuration spaces through the inverse scattering lifetime, and helps to clarify the role of the degrees of freedom in Strontium-doped Lanthanum Cuprate. Finally, we think that the self-consistent numerical procedure with the reduced phase space, induces nonlocality in the inverse scattering lifetime.

**Keywords:** Strontium-Doped Lanthanum Cuprate, Reduce Phase Space, Configuration Space, Phase Space, Zero Temperature Elastic Scattering Cross-Section, Lifetime, Mean Free Path, Numerical Modelling

## 1. Introduction

The field of high-temperature superconductivity started with the discovery by Bednorz and Müller in 1986 of the ceramic composite known as Barium-doped Lanthanum Cuprate (BaLaCuO system) with a transition temperature ($T_c$) of 30 K [1]. Experimental doping was an initial feature to obtain superconducting transition samples in ceramics with copper-oxygen planes. In this direction a few months later it was found another compound, the Strontium-doped $La_2CuO_4$ Cuprate with $T_c$ near 40 K, and this system was called Strontium-Doped Lanthanum Cuprate ($La_{2-x}Sr_xCuO_4$) and abbreviated as LaSrCuO [2]. The group of materials discovered by Bednorz and Müller was called the LaCuO family, composed by two high temperature superconducting ceramics (HTSC): The BaLaCuO, and the LaSrCuO systems. In 1987, another discovery in HTSC was made by Wu and collaborators in the compound Yttrium Barium Copper

Oxide (YBa$_2$Cu$_3$O$_{7-\delta}$). Wu et al. observed experimentally a transition from normal to superconducting state (NS) approximately at T$_c$ = 93 K, and it was called the YaBaCuO system [3]. Two features were inferred in 1986: First, the experimental HTSC values obtained for the T$_c$ in all these families, suggested difficulties to explain their microscopic behavior using the Bardeen, Cooper, and Schrieffer theory of superconductors (BCS) [4]. Second, all compounds classified as HTSC had a common feature, they were anisotropic layered structures containing copper-oxygen (CuO$_2$) planes [5-7].

The first experiments investigating the influence of impurities in HTSC showed that nonmagnetic disorder suppresses superconductivity [8] stronger than magnetic impurities, that destroy it in BCS superconductors [9]. Momono and Ido in 1994 performed an experimental work showing that the Strontium-doped Lanthanum Cuprate with increasing amounts of strontium doping (x) decreases the experimental value of T$_c$ in the research [10]. Particularly, it was contrasted superconducting low specific heat data at constant pressure (c$_p$) with the residual (N$_0$) density of states (DOS) using the relationship for the specific heat jumps ($\gamma$) with the relation $\gamma/\gamma_n \sim N_0/N_F$ where N$_F$ is the DOS at the Fermi level. In 1995, Sun and Maki in their work [11] used the theoretical formalism of the Larkin equation [12] $\ln \frac{T_c}{T_{c0}} = \psi\left(\frac{1}{2}\right) - \psi\left(\frac{1}{2} + \eta_c\right)$ with $\eta_c$ the pair breaking parameter, and T$_c$/T$_{c0}$ the dirty/clean superconducting critical temperature T$_c$. These works performed a fitting confirming that the experiments, suggested that Strontium-doped Lanthanum Cuprate is in the unitary elastic scattering regime [11]. This can be theoretically study using non-relativistic scattering theory as firstly suggested by Pethick and Pines in their research [13].

In the same order of experimental ideas, La$_{2-x}$Sr$_x$CuO$_4$ possesses a robust structure with stronger bonds than other HTSC ceramics, allowing larger crystals to be grown, and neutron scattering experiments can probe the material's magnetic structure with LSCO large bulk single-crystals [2]. Thus, another experiment worthy to mention here and related to the formalism of the ZTCS is the measurement of the universal limit for the electronic thermal conductivity ($\kappa_0$). This experiment helped to clarify the unconventional nature of the LSCO family, and also to study the insulator and superconducting phases in the universal limit as a function of disorder. This was one of the first experiments that sketched the phase diagram of the LSCO family [14]. Although the plot of the universal superconducting electronic thermal conductivity in this material $\kappa_{0s}(x)$ does not resemble the measurements of an unconventional bulk superconductor crystal as function of temperature, the experimental measurement of the universal limit as function of doping, showed phases with different material´s behavior (figure 2 in the research [14]).

It was found out that $\kappa_{0n}(x)$ (the normal state electronic thermal conductivity) is larger than $\kappa_{0s}(x)$ (the superconducting thermal conductivity) in the overdoped superconducting region [14], suggesting experimentally that in HTSC with a d-wave pairing, $\kappa_{0n}(x)$ should always be much larger than $\kappa_{0s}(x)$ in the superconducting state [15], thus remarking differences in the number of available quasiparticle heat carriers in these two phases. Hussey discussed several important experimental issues [16], such as, the Monomo et al. specific heat experiment [11] remarking that the downturn in c$_p$(T) at low temperatures was resolved in their overdoped samples, indicating a large T$^2$ term with a coefficient in agreement with a OP of a d-wave type. In addition, Hussey addressed an instructive discussion on the universal conductivity of several cuprates in the research [16], as the work by Sun and Maki [11], remarking that in their calculation, the universal $\kappa_{0s}(x)$ increases with doping. This variation in cuprates with a first order power in $n_{imp}$, implies that the creation of quasiparticles by the pair breaking nonmagnetic mechanism [12] is more important than the shortening of the dressed quasiparticle mean free path $l$ due to the impurity scattering. This issue is relevant to our work, since we make use of the inverse scattering lifetime as an output parameter.

Hussey also discussed from an experimental point of view, the nodal dressed quasiparticle spectrum of a d-wave superconductor, and the effects of impurities in normal and superconducting metallic alloys within the self-consistent T-matrix approximation [16], where nonmagnetic impurity scattering plays a dominant role in the transport, and thermodynamic properties of several unconventional superconductors due to the presence of a gap with line nodes. Finally, it is worthy to mention a recent experimental work where a summary of several techniques, including the angle-resolved photoemission spectroscopy (ARPES) are given [17]. Yamase et al., addressed the Compton scattering experiment, and suggested that it might reveal the Fermi surface structure in the underdoped region in cuprates if ARPES information is added, since these two experimental techniques are compatible [17]. From the theoretical point of view, it is worthy to mention that the tight-binding framework has been used recently within the Hubbard model [18], since we also use a tight-binding approach but within the ZTCS formalism. Finally, Walker developed a theoretical proposal of the Fermi liquid effects for anisotropic HTSC in his work [19], and for Fermi and Bose trapped atomic gases at ultra-cold temperatures, the importance of the ZTCS has been discussed as well by Pitaevskii in his research [20].

This work is summarized as follows. Section 1, introduces some experimental works of interest for the compound Strontium-doped Lanthanum Cuprate that points towards a unitary scattering regime scenario within the ZTCS. Section 2, addresses briefly some points of the ZTCS formalism in unconventional superconductors. Thus, it is explained the use of a singlet 1$^{st}$ harmonic OP with linear nodes including self-consistency, and the use of the Edwards nonmagnetic disorder scattering technique within a tight-binding (TB) framework, with the following parametrization: the zero temperature superconducting gap, the Fermi energy, the first neighbor's interaction, the strength of the elastic scattering, and the concentration of nonmagnetic impurities. Section 3 is devoted to the study of the variation of the zero temperature superconducting gap ($\Delta_0$) for La$_{2-x}$Sr$_x$CuO$_4$, contrasting some points with the values found for the compound Strontium Ruthenate [21].

Section 3 is considered the most important in this work and extensively discusses from a numerical perspective, two regions of importance for the Fermi averages, where the degrees of freedom of the dressed quasiparticles in the NS transition are separated depending on the value for the Fermi energy "$\varepsilon_F$", and shows how the self-consistent procedure with the ZTCS finds in the reduced phase space (RPS) the NS transition, when the frequency window is sufficiently large. Namely, there are found different zero gaps, namely, $\Delta_0$(input) and $\Delta_0$(output). Additionally, numerically the properties of $\tau^{-1}$ with a nonlocal self-consistent character, and the minimum value of the imaginary part of the ZTCS is found to be positive for a line nodes OP, with a geometrical sharply peak, contrasting that the separation of the degrees of freedom has two different physical mechanisms. Section 4 concludes, and section 5 outlines some recommendation for future works in this direction, comparing Strontium Ruthenate, and metallic normal thin films, with the layered behavior in Strontium-doped Lanthanum Cuprate.

## 2. The Zero Temperature Elastic Scattering Cross-Section (ZTCS)

### 2.1. The Role of the Inverse Self-Consistent Lifetime and the Mean Free Path in the Normal and Superconducting States

In this subsection, we recall the role of two fundamental physical parameters used in non-equilibrium statistical mechanics [22-24], and their relation with the nonrelativistic ZTCS when studying the NS with nonlocality expressed by the self-consistent equation [25]: the inverse scattering lifetime ($\tau^{-1}$), and the mean free path ($l$) which is the lattice constant parameter in the unitary collision regime. The function $\tau^l$ is calculated for Strontium-doped Lanthanum Cuprate considering five parameters in the equation for the ZTCS, when there are used rationalized Planck units (h/2$\pi$ = k$_B$ = c = 1). The ZTCS is given for the superconducting state by $\widetilde{\omega}(\omega + i\, 0^+)$ [26, 27].

$$\widetilde{\omega}_{sup}(\omega + i\, 0^+) = \omega + i\,\pi\,\Gamma^+ \frac{g(\widetilde{\omega})}{c^2 + g^2(\widetilde{\omega})}, \quad (1)$$

with the inverse strength parameter $c = (\pi\, N_F\, U_0)^{-1}$ where N$_F$ is the density of states at the Fermi surface and U$_0$ is the strontium impurity potential. The parameter $\Gamma^+ = n_{imp}/(\pi^2\, N_F)$ is proportional to the impurity concentration $n_{imp}$ in its first power [28]. The function $g(\widetilde{\omega})$ characterizes the superconducting state, strongly depends on Edwards disorder [29], and has a functional form strongly attached to the Fermi surface average with $g(\widetilde{\omega}) = \langle \frac{\widetilde{\omega}}{\sqrt{\widetilde{\omega}^2 - \Delta^2(k_x,k_y)}} \rangle_{FS}$, and where $\Delta(k_x, k_y)$ is the order parameter that contains the zero temperature energy gap ($\Delta_0$). In the normal state, $g(\widetilde{\omega}) = 1$, and the elastic scattering cross-section is given according to [30].

$$\widetilde{\omega}_{nor}(\omega + i\, 0^+) = \omega + i\,\pi\,\Gamma^+ \frac{1}{c^2 + 1}. \quad (2)$$

We are only interested in the unitary collision regime which applies if c << 1 and $l\,a^{-1} \sim 1$. Thus, (1) and (2) become respectively

$$\widetilde{\omega}_{sup}(\omega + i\, 0^+) = \omega + i\,\pi\,\Gamma^+ \frac{1}{g(\widetilde{\omega})} \quad (3)$$

$$\widetilde{\omega}_{nor}(\omega + i\, 0^+) = \omega + i\,\pi\,\Gamma^+ \quad (4)$$

The Fermi surface average $\langle ... \rangle_{FS}$ is performed in two limits of relevance for "$\xi(k_x, k_y)$" according a technique successfully used to fit experimental out of equilibrium data on the low temperature limit, i.e., the superconducting sound attenuation [31], the electronic thermal conductivity [32], and the electronic specific heat [33] of another unconventional superconductor with Strontium.

The mean free path "$l$" does not have an equation in the unitary limit, and can be considered an input parameter, since the relationship that holds is $l\,k_F \sim l\,a^{-1} \sim 1$, it means that the mean free path is equal to the lattice constant, and to the inverse $k_F^{-1}$ Fermi momentum (check Table 1). In this work, we consider only the unitary collision regime of the dressed fermion incoherent/coherent quasiparticles, with the level of impurity concentration denoted by the constant "$\Gamma^+$", and in addition, (1), (2), (3) and (4), that are in agreement with the Larkin equation for nonmagnetic disorder [12], supporting the experimental evidences discussed in the introduction. A recent numerical work that uses in addition to the unitary, other two scattering regimes, i.e., the hydrodynamic limit with $l >> a$, and the intermedia regime, where $l > a$, for which the inverse lifetime has a different behavior is given in the research [34]. The parametrization used for disorder ($c$, $\Gamma^+$) follows the numerical procedure introduced in the works [34, 35].

In this model, Strontium-doped Lanthanum Cuprate OP, possesses a linear nodal structure [15, 36], and it is represented by a scalar point group states with an even function of k, and a parity $i$ for the 1D irreducible representation B$_{1g}$ of the D$_{4h}$ point group. Table 1 summarizes the model used for computational purposes. The linear nodes of the 1D irreducible representation B$_{1g}$ show two features in the RPS: First, the nonlocality of $\tau^{-1}$. Second, the differentiation of the problem in two regions with different degrees of freedom, following the Gibbs distribution in nonequilibrium statistical mechanics [37].

*Table 1. Shows relevant input information used for the extensive numerical simulation.*

| Type of nodes | Fermionic dispersion | Order parameter | Scattering lifetime | Mean free path |
|---|---|---|---|---|
| Line nodes intersecting the Fermi surface represented by 4 pockets at the corners $+/-(\pi,\pi)$ of the 1st BZ (figures 2, 7) shadowed moccasin. | $\xi(k_x, k_y) = \epsilon_F + \xi_{hop}(k_x, k_y)$, depends on 2 parameters: The Fermi energy value and the hopping function. | Scalapino 1st harmonic line nodes model (figures 2 & 7) sketched in yellow color and with the singlet 1D irrep basis $\phi(k) = cos(k_x a) - cos(k_y a)$ | An output nonlocal self-consistent frequency dependent function & 5 simulation parameters in the unitary regime: $\Im[\widetilde{\omega}(\omega + i\, 0^+)] = 1/{2\tau_U}(\widetilde{\omega}(\omega), c = 0, \Gamma^+)$ | $\ell \sim a \sim$ constant is an input parameter |

We distinguish two equations where the imaginary ZTCS plays a fundamental role in nonequilibrium statistical mechanics following the Gibbs distribution in phase space [38], if the degrees of freedom of the physical system are taking into consideration. First, if the function $\Im[\widetilde{\omega}(\omega + i\, 0^+)] = (2\tau)^{-1}$ [39-40] is used to solve the $\tau$-approximation Boltzmann equation of a normal metal in the phase space with variables *(q,p)* for the nonequilibrium distribution of the dressed fermionic quasiparticles, where follows that $(\partial f/\partial t)_{coll} = -\tau^{-1}(f - f_0)$. If $f(t)$ goes rapidly to its equilibrium value $f_0$ and the inverse collision lifetime is approximated by a constant, such that $\tau^{-1} = 2\, \Im[\widetilde{\omega}(\omega + i\, 0^+)] = 2\pi\, \Gamma^+$ using (3-b).

The other situation, where the degrees of freedom are inferred from the imaginary part of the ZTCS is given by the time dependent probability density distribution of the configuration space [41], i.e., $\mathcal{W}(t) = |\psi_\omega(t)|^2 = \mathcal{W}_0 e^{-2\,\Gamma/\hbar\, t}$ with a wave function $\psi_\omega(t)$ that has an extra exponential imaginary damping "Γ" at the quasi-stationary level [42]. Thus, in this case $\mathcal{W}_0$ denotes an equilibrium situation, and if $\mathcal{W}(t)$ goes fast to its equilibrium value $\mathcal{W}_0$, we have that the following equation for $\mathcal{W}(t)$ holds $(\partial \mathcal{W}(t)/\partial t)_{qsd} = -2\,\Im[\widetilde{\omega}(\omega + i\, 0^+)]\,\mathcal{W}(t)$.

In both situations, the type of nonlocality, and the degrees of freedom come within in the inverse scattering lifetime described by collisions "*coll*" or a quasi-stationary damping "*qsd*" in the first partial derivatives with respect to time in the previous expressions. The imaginary function $\Im[\widetilde{\omega}(\omega + i\, 0^+)]$ expresses nonlocality, if the irreducible representation $B_{1g}$ with even parity follows an interplay between onsite and hopping terms in the Fermi average. It can be seen from the visualization of the RPS, and from the union of the phase space in nonequilibrium statistical mechanics, with the configuration space in nonrelativistic quantum mechanics (see figure 1 for an infographic summary of this section). Furthermore, since $\Im[\widetilde{\omega}(\omega + i\, 0^+)]$ at $T_c$ is always positive for all real input frequencies, and changes the sign of the slope in the imaginary self-consistent part at the point where the NS transition happens, it has a peculiar geometrical shape for singlet scalar pairing states, and never comes close to zero, contrasting with the case of strontium ruthenate which possesses triplet odd pairing, and has a small tiny gap [43] for the quasinodal case [21], within the TB disorder framework [44].

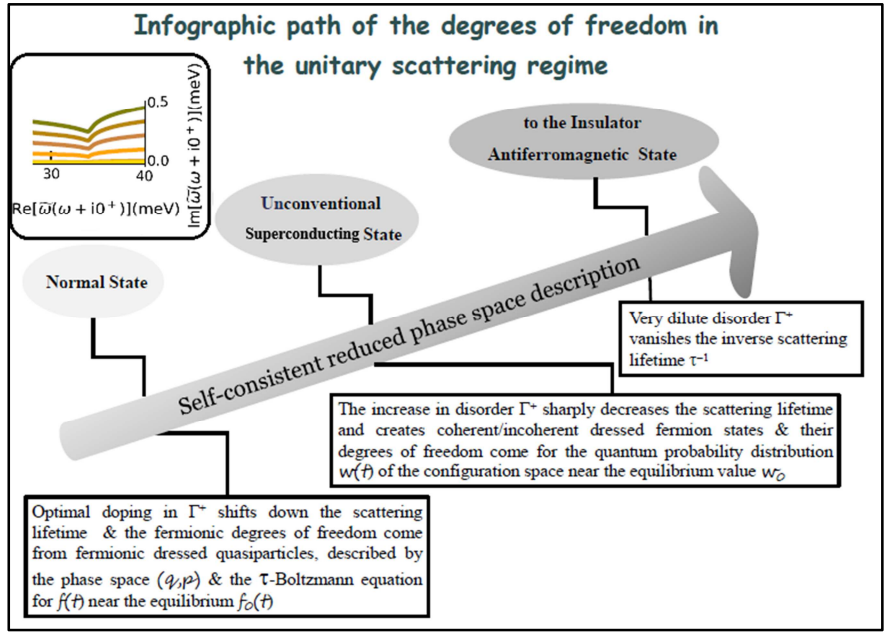

*Figure 1. Infographic path of the degrees of freedom in Strontium-doped Lanthanum Cuprate with an insert of the inverse lifetime for the $\epsilon_F$ = -0.4 meV case.*

This type of analysis was pointed out firstly in the work [37], where it was stated that not the entire microscopic motion of a physical system in statistical mechanics that follows Gibbs's distribution, the motion of the quasiparticles is quasi-classical, and happens only for some degrees of freedom. However, for the rest of degrees of freedom, the motion is quantized, and the energy containing those degrees of freedom can be written as function of energy with a quantum number n, i.e., "$E_n (q, p)$", that in our case corresponds to the description using the time-dependent probability density distribution with a negative damping given by $\gamma = -\Im[\tilde{\omega}(\omega + i\,0^+)]$ [38] and represents a classical window to quantum phenomena [45], but in this case, comes from the configuration space in nonrelativistic quantum mechanics [41].

### 2.2. The Role of the Fermionic Dressed Quasiparticles

The Fermi-Dirac distribution describes the behavior of the quasiparticles on the quasi-stationary quantum energy levels [25] ($\varepsilon_n$ and where n = 0,1,2...) with $f_n = (e^{-\frac{\varepsilon_n - \varepsilon_f}{k_B T}} + 1)^{-1}$. Therefore, it is important to recall that the Fermi energy $\varepsilon_F$, enters as a parameter in the Fermi-Dirac distribution $f_n$. It can be numerically controlled when the Fermi averages $\langle ... \rangle_{FS}$ are performed. In addition, the consequence of increasing the number of dressed negative fermion quasiparticles in the system results in an increase of the Fermi energy from negative to positive values [46]. Furthermore, in the doped ceramic $La_{2-x}Sr_xCuO_4$, the NS transition depends on both the concentration of doped ions, and the number of $CuO_2$ layers with a reservoir of fermionic quasiparticles with negative Fermi energy values. Check figure 2 (b) in the research [7], where it has been proposed that the partial substitution of Strontium for Lanthanum in a solid solution of $La_{2-x}Sr_xCuO_4$, obeys a manipulation of the charge reservoir, i.e., the change of one electron per copper site, making the copper valence higher than $Cu^{+2}$, and inducing unconventional superconductivity. Cava, also points out that in the antiferromagnetic phase with the insulator $La_2CuO_4$, the $Cu^{+2}$ "½" spin ions (with one unpaired electron per copper) are located in the $d_{x2-y2}$ atomic orbitals [7].

In order to show the link with the degrees of freedom mechanism and nonlocality in this case, small colored insertion in figure 1 shows the behavior of the function $\Im[\tilde{\omega}(\omega + i\,0^+)]$ as function of $\omega = \Re[\tilde{\omega}(\omega + i\,0^+)]$ with parameters: $\omega = \Delta_0 = 33.9$ meV, $\varepsilon_F = -0.4$ meV, $c = 0$, $t = -0.2$ meV and $\Gamma^+ = (0.005, 0.01, 0.05, 0.10, 0.15, 0.20)$ meV, indicating an impurity concentration varying from very dilute disorder with $\Gamma^+ = 0.005$ meV, forming a robust coalescent metallic phase with constant scattering lifetime; to an optimal disorder phase with very incoherent fermionic quasiparticles in both normal, and superconducting states. Remarkably, figure 1, shows for the range of doping $\Gamma^+ = 0.05 - 0.20$ meV a noticeable sharp peak with the same geometrical shape, with a change in slope in the imaginary ZTSC around the real frequency value of $\omega = 33.9$ meV $= \Delta_0$; that can be attributed to the change in the type of degrees of freedom at the NS phase transition.

The ZTCS for Strontium-doped Lanthanum Cuprate is more difficult to simulate than for Strontium Ruthenate ($\Delta_0 = 1$ meV), because the real frequency window should suffix to locate NS transition point; and in addition; we cannot extend this procedure to the antiferromagnetic phase. This is due to the existence of gap values that strongly depend on Edwards type of disorder [29], and the numerical calculation depends on the Fermi energy, with real frequencies in a different range, needed to describe the RPS behavior, and the types of degrees of freedom. In the normal, and superconducting states, the degrees of freedoms change in nature since superconducting pair excitations appear, although not equally in Strontium-doped Lanthanum Cuprate, and Strontium Ruthenate [47, 48]. We know, as suggested in the previous subsection, that self-consistency is unavoidable, and a good dispersion relation for the reduced phase space is given by $\omega\,\tau(\tilde{\omega}(\omega)) \sim 1$ with $\omega \sim 1/\tau \sim 4\,\Delta_0$. This suggest a frequency window of $\pm 120$ meV to find the NS transition for Strontium-doped Lanthanum Cuprate, as we are able to predict numerically in section 3.

### 2.3. Pairing Singlet Scalar States from the 1D Irreducible Representation $B_{1g}$ of the Point Group $D_{4h}$

We use for simulation (check Table 1) the 1D irrep. $B_{1g}$ for the tetragonal point group $D_{4h}$, in a square lattice with the 1st harmonic given by $\phi(k_x, k_y) = cos(k_x a) - cos(k_y a)$ which has an even parity $\phi(k_x, k_y) = \phi(-k_x, -k_y)$ [15, 36]. The OP is a scalar function, that has dependence of the zero temperature gap ($\Delta_0$) near the line nodes, and is given by $\Delta(k_x, k_y) = \Delta_0\,\phi(k_x, k_y) = \Delta_0[cos(k_x a) - cos(k_y a)]$. However in the literature is also found the value $\Delta_0/2$ for the function $\Delta(k_x, k_y)$ [49], so we would like to clarify this point before presenting the numerical results.

Yoshida et al., performed detailed angle-resolved photoemission spectroscopy experiments (ARPES) in several cuprates including Strontium-doped Lanthanum Cuprate [49]. They were able to explain the difference between two magnitudes: the gap near the nodes ($\Delta_0$) and the antinodal gap ($\Delta^*$). They pointed out that $\Delta^*$ is approximately independent, if doping is fixed for the material parameters. Nevertheless, Yoshida et al. suggested that $\Delta_0$ strongly depends on the material properties such as doping, and is able to track the magnitude of $T_c^{max}$. Henceforth, we are able to show in the next section, that using the ZTCS, we can track as well the zero superconducting gap behavior numerically, differentiating two regimes. One where dressed holes dominate the physics with $\varepsilon_F$ negative, and far from the zero value. The other case represents an increase of the amount of fermionic electrons, and $\varepsilon_F$ closer to the zero value. In addition, it is important to mention that figure 3 in the work [49] shows what Yoshida et al. called, the leading edge midpoints gap (LEM) with $\Delta_{LEM} = \Delta_0/2$.

## 3. Numerical Behavior of the Zero Temperature Superconducting Gap

### 3.1. General Considerations

The determination of the crossover between the normal, and superconducting phases (NS) in HTSC cuprates including the ceramic Strontium-doped Lanthanum Cuprate is still one of the main task, from the experimental, theoretical and numerical points of view. It continues to be a matter of an intense debate, despite this compound is one among the first discovered cuprates. The experimental presence of fermion incoherent carriers in its transport properties for different doping levels, and the localization of the NS transition temperature using angle-resolved photoemission spectroscopy experiments, help to clarify several points, such as, the presence of different types of superconducting gaps [49]. We think that to accomplish a unique theory to describe the whole phase diagram in cuprates is quite difficult because studying the ZTSC, we have observed how disorder (nonmagnetic dirt) numerically affects, the 2D coherence of dressed fermionic quasiparticles in each single layer, when five parameters are considered within a tight-binding self-consistent framework. Thus, the results presented in the next subsections, aim at clarifying this point, from a numerical perspective.

We have noticed, that using the ZTSC formalism in the unitary regime, there are not small parameters in the solution of the self-consistent scattering lifetime for an OP with linear nodes. The minimum observed at the NS transition is finite and greater than cero, preserving its geometrical shape in all cases, except for a small positive value when the formalism with a very dilute doping is numerically solved, as seen in the insert of figure 1 ($\Gamma^+$ = 0.005 meV). Disorder only enters in first power of $n_{imp}$ in the self-consistent ZTSC formalism [28]. Thus, we clarify this point checking the numerical robustness, with a well-established numerical self-consistent ZTSC procedure already tested by us in several cases of relevance. Furthermore, we have observed two different physical behaviors when solving this numerical problem. One when the parameter that represents the Fermi energy is $\varepsilon_F$ = - 0.4 meV; and another behavior, when $\varepsilon_F$ = - 0.04 meV. Henceforth, we proceed to vary the value of the zero superconducting energy gap for two values of the Fermi energy, and to observe, how the crossover of the NS transition occurs.

### 3.2. Zero Superconducting Gap Variation of the Scattering Cross-Section for Curved Pockets

The 1st simulation (sim) performed in this subsection, refers to a set of parameters corresponding to the nearest neighbor TB expression for a single band centered at the corners (±π/a, ±π/a) of the 1st Brillouin 2D zone (figure 2), with a dispersion relation $\xi(k_x, k_y) = \epsilon_F + \xi_{hop}(k_x, k_y)$ where the quasi-momentum hopping dependent part of the dispersion is given by $\xi_{hop}(k_x, k_y) = 2\, t\, [\cos(k_x\, a) + \cos(k_y\, a)]$, it is an even function $\xi_{hop}(k_x, k_y) = \xi_{hop}(-k_x, -k_y)$ of the quasi-momentum, and the numerical values of the tight-binding parameters are chosen as $(t, \epsilon_F)$ = (- 0.2,- 0.4) meV with the zero gap given by the yellow crossing lines (figure 1).

Figure 2 shows the implicit sketch for the expression, $0 = -0.4 + \xi_{hop}(k_x, k_y)$, and the numerical values given in the previous paragraph. The pockets shadowed moccasin color are centered symmetrically at points $\pm(k_x, k_y) = \pm(\pi, \pi)$. The shape of the pockets situated in the corner are part of a circle. The yellow crossing lines represent the implicit sketch of the linear OP nodes, for the 1D irrep $B_{1g}$. The scattering strength is the unitary collision limit with c = 0. The doping $\Gamma^+$ is a discrete variable of two values. One represents a diluted doping $\Gamma^+$ = 0.05 meV, and the second simulates optimal doping value with $\Gamma^+$ = 0.20 meV, already tested for a small frequency window, for three scattering regimes [34]. Using these parameters, we vary, the value of the zero temperature energy gap in the imaginary part of the ZTCS in (3a) and (3b). We use four values, namely, the cero value for the normal state and three values for the superconducting phase. In the first simulation shown in figure 3, we used the values $\Delta_0$ = (0.0, 20.0, 33.90, 60) meV with a dilute doping dependence $\Gamma^+$ = 0.05 meV and $(t, \epsilon_F)$ = (-0.2,- 0.4) meV.

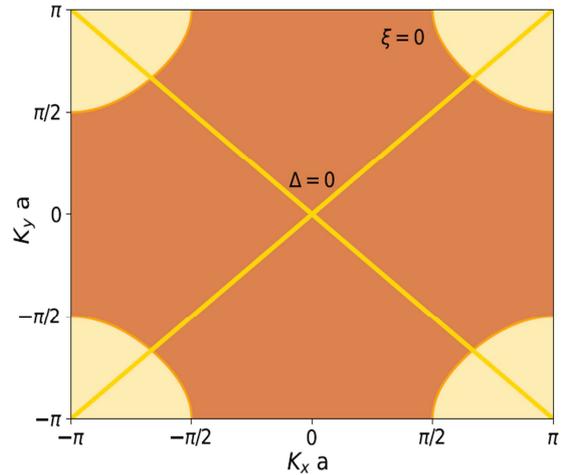

*Figure 2.* Implicit fermionic dispersion (shadowed moccasin) and bosonic excitation pairs (shadowed yellow) for $\varepsilon_F$ = -0.4 meV.

The value $\Delta_0$ = 0.0 meV indicates the normal state and it is given by (3b). The values $\Delta_0$ = (20.0, 33.90, 60.0) meV show the superconducting state behavior of the imaginary part of the ZTCS given by (3a). The diluted doping $\Gamma^+$ = 0.05 meV has a resemblance of a coalescent metallic phase in Strontium-doped Lanthanum. The results of the simulation using a frequency window with limiting values at real frequency points equal to $\pm\, 80.0\, meV$ in the RPS, are shown in figure 3.

These results indicate that the input zero superconducting energy gap parameter for the four values has the following characteristics in the RPS:

1) The unitary peak in the imaginary function is increased with the input of higher values of $\Delta_0$ from the constant value of 0.16 meV in the normal state to almost 1.75 meV for the maximum $\Delta_0$ = 60.00 meV.
2) The normal state of Strontium-doped Lanthanum Cuprate has a cero slope line, indicating a constant imaginary value, and therefore a constant unitary scattering lifetime, when the doping is fixed. Its value is approximately 0.16 meV, and it agrees with (3b).
3) The transition from the normal to the superconducting phase (NS) remains fixed as having the same input and output values for the zero energy superconducting gap $\Delta_0$ = (20.0, 33.90, 60.0/60.01) meV, when the Fermi energy parameter is $\varepsilon_F$ = - 0.4 meV. Thus, in this case $\Delta_0$, and also, the degrees of freedom are not sensitive to the material parameters.

4) The minimum in the imaginary part of the reduced phase space, i.e., the function $\Im[\widetilde{\omega}(\omega+i\,0^+)]$ has two features:
   a) First, an almost equal finite value for the zero energy superconducting gap values equal to (20.0, 33.90, 60.0) meV is given respectively by the values (0.06, 0.06, 0.05) meV.
   b) Second, there is a sharp minimum with a similar geometrical shape at the three NS points.

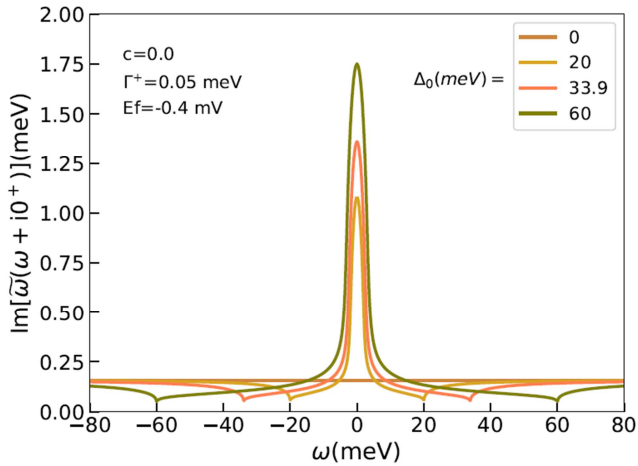

**Figure 3.** *1$^{st}$ RPS simulation with a Fermi energy $\varepsilon_F$ = -0.4 meV for 4 superconducting gap values & diluted doping in a ±80 meV frequencies range.*

If the 2$^{nd}$ sim is performed, with real frequencies in an interval [-100, +100] meV, the plots are shown in figure 4, and indicate that a bigger real part of the reduced phase space, $\omega = \Re[\widetilde{\omega}(\omega+i\,0^+)]$, but it does not introduce new results.

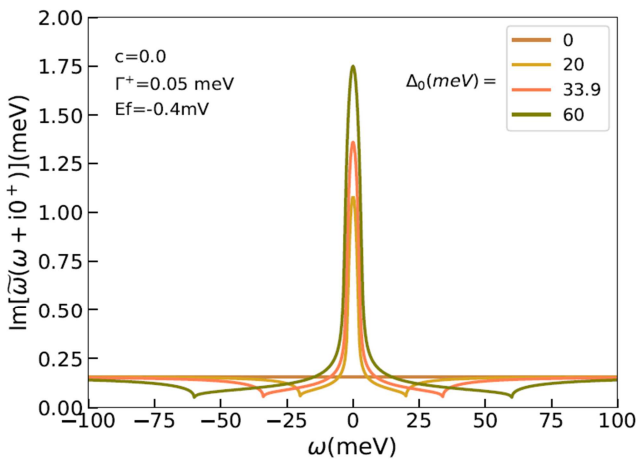

**Figure 4.** *2$^{nd}$ RPS simulation with a Fermi energy $\varepsilon_F$ = -0.4 meV for 4 superconducting gap values & diluted doping in a ±100 meV frequencies range.*

Subsequently, we perform the 3$^{rd}$ simulation in the RPS with an optimal impurity concentration value, i.e., $\Gamma^+ = 0.20$ meV. The graphical results are shown in figure 5.

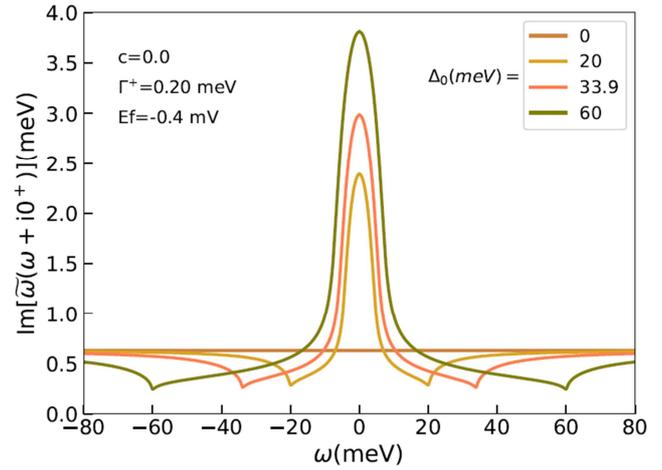

**Figure 5.** *3$^{rd}$ RPS simulation with a Fermi energy $\varepsilon_F$ = -0.4 meV for 4 superconducting gap values & optimal doping in a ±80 meV frequencies range.*

The results from figure 5 indicate that, the input zero superconducting energy gap parameter for the three values, has almost the same output value (all gap values decrease for less than 1 %). However, new characteristics in the reduced face space, compared to figure 3 are encountered:

1) The unitary peak in the imaginary function is increased with higher values of the $\Delta_0$ superconducting parameter from 0.63 meV in the normal state to almost 3.80 meV for $\Delta_0$(output) = 59.95 meV. So, it represents a smaller scattering lifetime than for the case with a dilute amount of impurities, i.e., dressed quasiparticles scatter with more frequency, due to an optimal doping.
2) The normal state is a zero slope line, indicating a constant imaginary value, and therefore, a constant inverse unitary scattering lifetime, when the doping is fixed. Its value, now is increased with respect to the diluted doping case, and is 0.63 meV. The same situation happens in the superconducting phase. Therefore, normal incoherent dressed states scatter with more frequency.
3) The NS transition remains almost fixed at the same input values for the zero energy superconducting gap, with $\Delta_0$ (output) = (19.90, 33.81, 59.95) meV, when the Fermi energy $\varepsilon_F$ = - 0.4 meV, pointing towards a superconducting gap simulation independent of the material parameters.
4) The minimum in the imaginary function has almost the same finite value for the zero energy gap, and it is given accordingly to (0.29, 0.26, 0.24) meV, indicating a very small decreasing.

Let us, for illustrative purposes, perform an identical material parameters 4$^{th}$ sim opening the RPS real frequency window to an interval of $\pm 100.0\,meV$. The plots are shown in figure 6, and clearly show that they do not introduce new numerical results, with respect to figure 5.

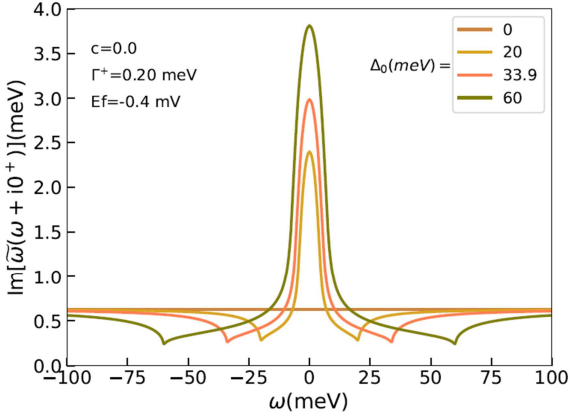

**Figure 6.** *4th RPS simulation with a Fermi energy ε$_F$ = -0.4 meV for 4 superconducting gap values & optimal doping in a ±100 meV frequencies range.*

Henceforth, the main numerical results from section 3.2 are summarized as follows: The position in the reduced phase space from the normal to the superconducting transition (NS) remains almost at the same zero temperature energy gap input value (within a numerical difference rounding 1% between input and output values), and therefore, it is easier to determine the real frequency needed for the self-consistent simulation. This happen only for values of the Fermi energy, where hold the following conditions: The absolute Fermi energy value |ε$_F$| ≈ 2 |t|, and the ε$_F$ value, it is found to be negative, and geometrically happens for circular shapes at the four corners of the 1$^{st}$ Brillouin zone.

It is seen from the simulation, that when the study is performed, the NS transition, and thus, the Δ$_0$ value in the RPS does not depend on the material parameters, and there is a single frequency point, where the degrees of freedom are separated into two groups: Those of the nonequilibrium statistical mechanics phase space, corresponding to the normal state, and the ones from the configuration space of nonrelativistic quantum mechanics corresponding to the superconducting phase, remaining, the zero gap input and output values, within 1% of numerical difference if ε$_F$ = -0.4 meV. This case corresponds to the infographic chart given by figure 1, and its reduced phase space insert in the left upper side.

### 3.3. Zero Superconducting Gap Variation of the Scattering Cross-Section for Quasi-Flat Pockets

The simulation performed in this subsection, refers to a set of nearest neighbor parameters for a single band with almost flat sheets centered at the corners (±π/a, ±π/a) of the 1$^{st}$ Brillouin 2D cell, shadowed moccasin color in figure 7, and with an identical expression, as for the quasi-circular pockets case, i.e., $\xi(k_x, k_y) = \epsilon_F + \xi_{hop}(k_x, k_y)$, where the hopping term is $2t\left[\cos(k_x a) + \cos(k_y a)\right]$, but now the numerical values of the TB parameters are the following ones: We keep the same hopping coefficient *t* = -0.2 meV, but vary the independent Fermi energy coefficient, to be near the zero value, i.e., ε$_F$ = -0.04 meV. This almost represent, the electron-hole symmetry case, and it is found, nearby the half-filling ground state, where interatomic quantum mechanical hopping is the dominant physical mechanism.

The implicit plot given according to $0 = -0.04 + \xi_{hop}(k_x, k_y)$ being $\epsilon_F = -0.04$ meV (|ε$_F$| << 2 |t|) resembles a geometrical ongoing transition to the half-filled antiferromagnetic ground state, and a geometrical flattening of the four circular pockets, centered symmetrically at points $\pm(k_x, k_y) = \pm(\pi, \pi)$. The main signature in the dispersion $\xi(k_x, k_y)$ is that for $\epsilon_F = -0.04$ meV, the dominant behavior comes from the quantum mechanical nearest hopping term $\xi_{hop}(k_x, k_y)$, closer to the insulator La$_2$CuO$_4$, as can be seen from figure 7. If $\epsilon_F \to \pm 0$ meV the four pockets will be joined together as four straight lines at the symmetric points $\pm(k_x, k_y) = \pm(\pi/2, \pi/2)$.

In such a hypothetical case, the average over the Fermi surface converts the 2D integration in (3a) into a line integral, and complicates the numerical procedure. Therefore, the value $\epsilon_F = -0.04$ meV suffixes our duty, allowing numerically, to compute a new behavior of the imaginary ZTCS. Additionally, the yellow lines represent, the implicit sketch of the linear OP nodes, i.e., $\Delta(k_x, k_y) = 0$.

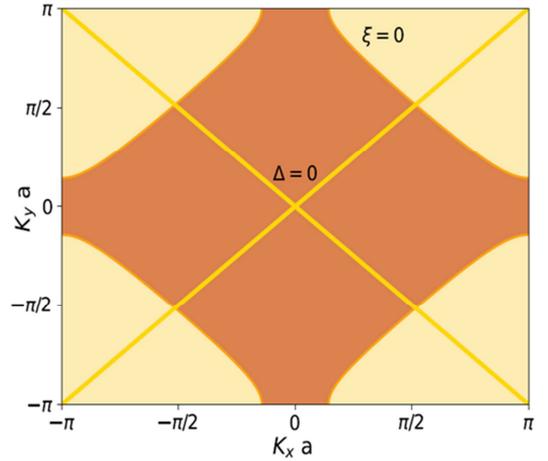

**Figure 7.** *Implicit fermionic dispersion (shadowed moccasin) and bosonic excitation pairs (shadowed yellow) for ε$_F$ = -0.04 meV.*

If the Fermi energy is $\epsilon_F = -0.04$ meV, the scattering strength used for the new simulation is given by the same expression in the unitary scattering limit (c = 0). The doping Γ$^+$ in this subsection has the same previous two values. A diluted doping Γ$^+$ = 0.05 meV, and an optimal doping value Γ$^+$ = 0.20 meV. Using these parameters, we change the values of the zero temperature energy gap in the imaginary part of the ZTCS within same range. The value Δ$_0$ = 0.0 meV, indicates the normal state given by (3b). The input zero gap values are the same four, i.e., Δ$_0$ = (20.0, 33.90, 60.0) meV.

The results of the 5$^{th}$ self-consistent sim, using a real frequency window between $\pm 80.0\ meV$ are shown in figure 8. In order to analyze the new behavior, let us check, the green plot for Δ$_0$ = 60.00 meV. The NS transition frequency point disappeared from the plot. There is not a sharply minimum indicating the separation of the degrees of freedom for the highest value of the gap.

Henceforth, figure 8 should be contrasted with figure 3 with the same value of the diluted doping parameter. We just have encountered a new remarkable numerical property of the self-consistent solution. The zero temperature superconducting energy gap values obtained as output values in the numerical self-consistent procedure are strongly distorted from the original input $\Delta_0$ values, accordingly to the following table:

***Table 2.** Partial filled table of the changes in $\Delta_o$ (Input) & $\Delta_o$ (Output) for $\varepsilon_F$ = -0.04 meV & dilute doping $\Gamma^+$ = 0.05 meV.*

| Zero sup. energy gap with dilute doping ($\Gamma^+$= 0.05 meV) | $\Delta_0$ normal state (meV) | $\Delta_0$ sup. state (meV) | $\Delta_0$ sup. state (meV) | $\Delta_0$ sup. state (meV) |
|---|---|---|---|---|
| Input values | 0.00 | ± 20.00 | ± 33.90 | ± 60.00 |
| Output values | 0.00 | ± 38.00 | ± 64.40 | It is not shown |
| \|$\Delta_o$ (Output) – $\Delta_o$ (Input)\| | 0.00 | 18.00 | 30.50 | --- |

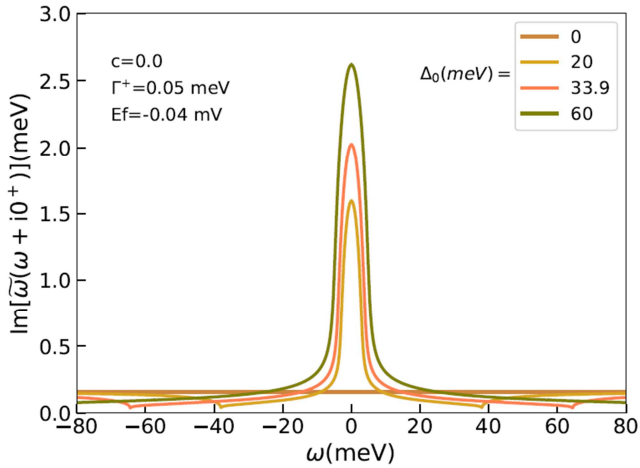

**Figure 8.** *5th RPS simulation with a Fermi energy $\varepsilon_F$ = -0.04 meV for 4 superconducting gap values & diluted doping in a ±80.00 meV frequencies range.*

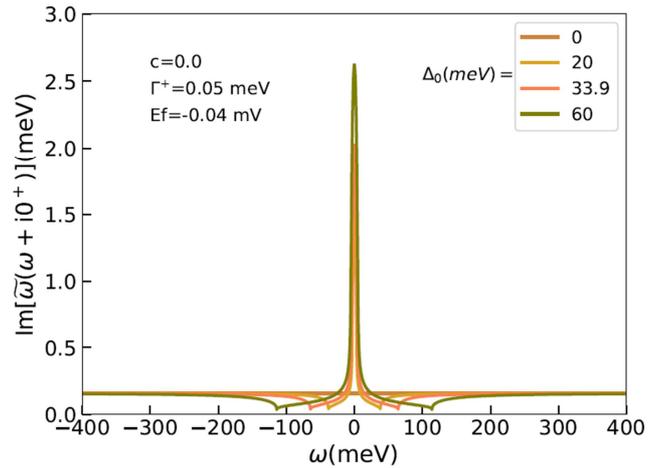

**Figure 9.** *6th RPS simulation with a Fermi energy $\varepsilon_F$ = -0.04 meV for 4 superconducting gap values & diluted doping in a ±400.00 meV frequencies range.*

Therefore, we see from figure 8, and table 2, that for a dilute doping of $\Gamma^+$ = 0.05 meV, there are strong differences between the input and output zero temperature gap values, and the value | $\Delta_o$ (Output) - $\Delta_o$ (Input) | is a nonzero value. This suggest that for the case when the Fermi energy is closer to zero, the zero energy gap is very sensitive.

This parameter distortion, additionally suggests, that the hopping term (it has an interatomic site character, probably with ions interaction) is dominant for an almost symmetric dispersion, when the area shadowed moccasin color increases, and changes the shape of the Fermi surface to an almost flat line, intercepting the 1st Brillouin zone closer to the symmetry points given by $\pm(k_x, k_y) = \pm(\pi/2, \pi/2)$. On the other hand, from the numerical point of view, if $\epsilon_F \to \pm 0$ meV, the calculation of the Fermi average of the ZTCS, becomes a numerical challenge, and the family of self-consistent solutions in the RPS are difficult to obtain using a self-consistent procedure based on a Fermi average integration, so a description of the insulator state in $La_2CuO_4$, we consider that is not possible within this numerical framework.

If the 6th simulation is performed using same material parameters in an interval [−400.00, +400.00] meV, the results are shown in figure 9, and clearly, we see the four output $\Delta_0$ values, with a dilute coalescent doping $\Gamma^+$ = 0.05 meV, remarkably showing the evolution of the zero energy superconducting gap starting from the normal state.

Therefore, all output $\Delta_0$ values for dilute doping are given in table 3. It shows numerically, how the zero energy superconducting gap values are displaced for an almost zero Fermi energy parameter (an order of magnitude smaller compared to the case given in figures 3-6). Of course, the numerical calculation does not indicate what $\Delta_0$ experimentally values for $La_{2-x}Sr_xCuO_4$ are the most appropriate. But the reduced phase space procedure using the self-consistent elastic scattering cross-section, remarkably shows for what material parameters, the value of $\Delta_0$ and therefore, the values of the transition temperature, to the superconducting state are sensible to the rest of the material parameters.

In another order of ideas, the *self-consistent reduced phase space*, using the set of five simulation parameters ($\varepsilon_F$, t, $\Delta_0$, $\Gamma^+$, c) is able to answer when the degrees of freedom given by the *phase space* in nonequilibrium statistical mechanics, and the *configuration space* in nonrelativistic quantum mechanics are more sensitive to material parameters in the case of having nonmagnetic doping with strontium in the NS phase transitions, as we have numerically demonstrated in this work. The physics corresponds to a quantum mechanical quasi stationary level case, and remains hidden in the Gibbs distribution [37].

Table 3. *Filled table of the changes in Δ$_o$ (Input) & Δ$_o$ (Output) for ε$_F$ = -0.04 meV & dilute doping Γ+= 0.05 meV.*

| Zero sup. energy gap with dilute doping (Γ$^+$= 0.05 meV) | Normal state (meV) | Δ$_0$ sup. state (meV) | Δ$_0$ sup. state (meV) | Δ$_0$ sup. state (meV) |
|---|---|---|---|---|
| Input values | 0.00 | ± 20.00 | ± 33.90 | ± 60.00 |
| Output values | 0.00 | ± 38.00 | ± 64.40 | ± 114.00 |
| ∣Δ$_o$ (Output) − Δ$_o$ (Input)∣ | 0.00 | 18.00 | 30.50 | 54.00 |

Other similitudes and differences, noticeable from figures 3 and 7-8 with dilute doping are the following:

1) The unitary peak in the imaginary function is increased for higher values of the Δ$_0$, from 0.16 meV in the normal state to 2.61 meV for the maximum gap, i.e., Δ$_0$ = 60.0 meV, and $\epsilon_F = -0.04$ meV. This shows that in the case, when the Fermi energy is closer to a cero value, the scattering lifetime is smaller than when $\epsilon_F = -0.4$ meV for a dilute doping (Γ$^+$ = 0.05 meV).
2) The normal state in the reduced phase space is represented by a cero slope line, as in the previous case, and the imaginary elastic scattering term has a constant value of 0.16 meV, so $\epsilon_F$ does no change the normal state unitary lifetime.
3) The minimum in the imaginary part of the reduced phase space has almost the same finite value for the new zero energy gap Δ$_0$ (output)= (38.00, 64.40, 114.00) meV, and it is given approximately by the value $\Im[\tilde{\omega}(\omega + i\,0^+)]$ = 0.04 meV (it is almost the same value obtained from figure 3).

We continue with a 7$^{th}$ simulation, changing the impurity concentration value to an optimal doping of Γ$^+$ = 0.20 meV. The sim is performed with the same set of input values Δ$_0$(input) = (20.0, 33.90, 60.0) meV, and the graphical results in the RPS are shown in figure 10, for a real frequency window ω = (−80.00, +80.00) meV.

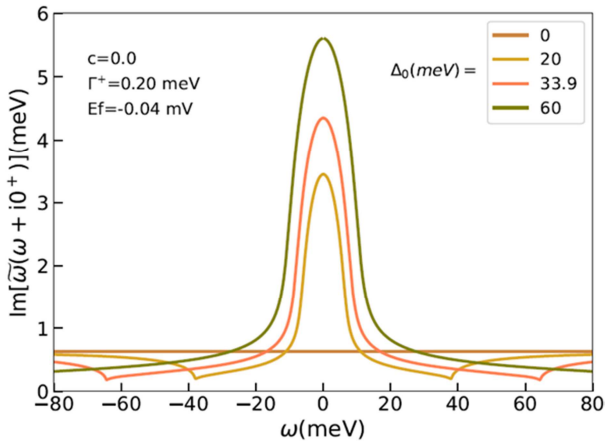

**Figure 10.** *7$^{th}$ RPS simulation with a Fermi energy ε$_F$ = -0.04 meV for 4 superconducting gap values & diluted doping in a ±80.00 meV frequency range.*

The plot in figure 10 should be compared with the plot given in figure 5. We see that the same effect of distorted values of the zero temperature gap, numerically appears for the Δ$_0$(output). Additionally, as in the previous dilute doping case, in the plot for the value Δ$_0$ = 60.00 meV, the NS transition is out of the frequency range, that is used in the 7$^{th}$ simulation. Since in this case, there are more fermionic dressed negative electrons than positive holes [25], and a Fermi energy of ε$_F$ = -0.04 meV, the scattering lifetime, which is inversely proportional to the imaginary part of the ZTCS is $(2\tau)^{-1} \sim 5.5$ meV, and $\tau$ becomes smaller that for the case of an almost circular pocket Fermi surface. We think that more dressed electrons are elastic scattered due to a higher amount of doping (Γ$^+$ = 0.20 meV), and this is, mainly due to the change in behavior of the Fermi-Dirac $f_n$ distribution function, as was noticed in section 2 [46], increasing the number of dressed negative fermion quasiparticles (electrons) in the system when the Fermi energy changes from negative to positive values, i.e., within the interval -0.4 meV ≤ ε$_F$ ≤ -0.04 meV.

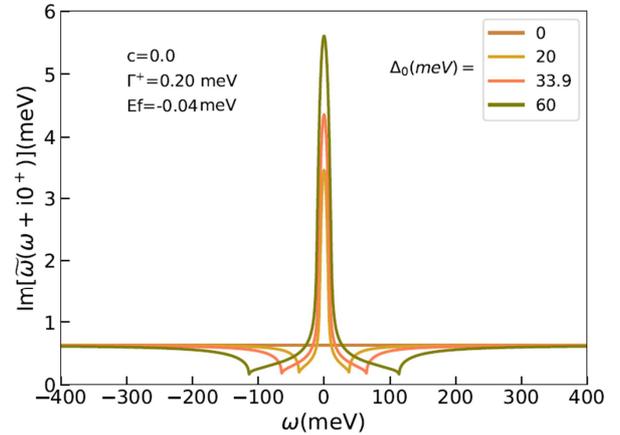

**Figure 11.** *8$^{th}$ RPS simulation with a Fermi energy ε$_F$ = -0.04 meV for 4 superconducting gap values & optimal doping in a ±400.00 meV frequency range.*

The final plot comes from the 8$^{th}$ simulation performed using identical material parameters, and a real frequency in the interval [−400.00, +400.00] meV. The results are shown in figure 11, where the output numerical values for the zero superconducting gap Δ$_0$ are obtained from the data of the corresponding plots, and a new Table can be filled out, with new output microscopic gap parameters.

In accordance with the previous statement, all output zero energy superconducting Δ$_0$ values are given in table 4. Figure 11 shows, how the zero energy superconducting gap values are displaced from input values, when it is used an small amount of the Fermi energy ($\epsilon_F = -0.04$ meV) for an optimal fixed doping value. The reduced phase space elastic cross-section procedure differentiates in this case, the material parameters needed to find the correct position of Δ$_0$ values, and again, we think that for optimal doping, the values of T$_c$ are quite sensitive to the rest of the input material parameters, depending on the different quasiparticle types of degrees of freedom.

**Table 4.** *Filled table of the changes in $\Delta_o$ (Input) & $\Delta_o$ (Output) for $\varepsilon_F$ = -0.04 meV & optimal doping $\Gamma^+$ = 0.20 meV.*

| Zero sup. energy gap with dilute doping ($\Gamma^+$= 0.05 meV) | Normal state (meV) | $\Delta_0$ sup. state (meV) | $\Delta_0$ sup. state (meV) | $\Delta_0$ sup. state (meV) |
|---|---|---|---|---|
| Input values | 0.00 | ± 20.00 | ± 33.90 | ± 60.00 |
| Output values | 0.00 | ± 37.90 | ± 64.30 | ± 113.90 |
| $|\Delta_o$ (Output) $-\Delta_o$ (Input)$|$ | 0.00 | 17.90 | 30.40 | 53.90 |

Other differences noticeable from figures 5, and 10-11, at optimal doping are the following:

1) The unitary peak in the imaginary function is increased with higher values of $\Delta_0$, from a normal state imaginary output of 0.63 meV to 5.60 meV for $\Delta_0$ (output) = 64.0 meV and $\varepsilon_F$ = -0.04 meV. This shows, that in the case when the Fermi energies are closer to a cero value, the scattering lifetime is smaller than when $\varepsilon_F$ = -0.4 meV for an optimal doping ($\Gamma^+$ = 0.20 meV).

2) The normal state is a zero slope line as in the previous case, and the imaginary elastic scattering term has a value of 0.63 meV, indicating that in the unitary limit, optimal doping, gives the same constant lifetime as for dilute doping, and $\varepsilon_F$ = -0.04 meV.

3) The minimum in the imaginary part of the RPS has almost the same finite value for the $\Delta_0$ (output) = (38.00, 64.40, 114.00) meV, and it is given by the values: $\Im [\widetilde{\omega}(\omega + i\, 0^+)]$ = (0.20, 0.19, 0.17) meV.

## 4. Conclusions

The present work was aimed at numerically investigating the behavior of the Fermi energy parameter ($\varepsilon_F$) in a singlet 2D layer of Strontium-doped Lanthanum Cuprate, using a self-consistent numerical zero temperature elastic scattering cross-section procedure in the unitary collision regime, varying the values of the zero temperature superconducting energy gap ($\Delta_0$). We found that there are two different reduced phase space tight-binding regimes:

1) One picture, where $\varepsilon_F$ = -0.4 meV indicates an asymmetric dressed quasiparticles scenario (holes), where material parameters and degrees of freedom of the NS transition are no sensitive to the numerical change on the microscopic parameter $\Delta_0$ in the unitary limit with a constant mean free path $l \sim a$. This scenario considers that the physical mechanism is induced by the parameters ($t, \varepsilon_F$), that have the same order of magnitude.

2) A different picture, where $\varepsilon_F$ = -0.04 meV indicates a dressed electron almost symmetric quasiparticles scenario, and where the material parameters are strongly sensitive to the numerical change of $\Delta_0$ and the degrees of freedom. The unitary limit with $l \sim a$ still prevails. The 2$^{nd}$ picture considers that the physical mechanism is induced by the small value of the Fermi energy parameter $\varepsilon_F$ = -0.04 meV with $|t| > |\varepsilon_F|$, and this picture tends to be related to the hopping nonrelativistic quantum mechanical degrees of freedom at the quasi-stationary quantum mechanical level.

3) We think that self-consistent nonlocality is an intrinsic property of the reduced phase space, where holds that $\omega\,\tau(\widetilde{\omega}) \sim 1$, and that in order to check for the $\varepsilon_F$ scenario that takes place, we need also to consider in the reduce phase space, real frequencies of the order of $\omega \sim 4\,\Delta_0$. To compare with other works, nonlocality can be also found for the mean free path "$l$" in thin metallic films when considering the anomalous skin effect, using Fermi averages for complicated Fermi surfaces. Theoretical details for the mean free path within the anomalous skin effect, can be found in the works [50, 51], and recent experiments are given in the articles [52-54].

The numerical parameter $\varepsilon_F$ is defined in several ways. If it is considered a nonrelativistic quantum mechanics study in the configuration space, with the fermionic quasiparticles dispersion $\xi(k_x, k_y) = \epsilon_F + \xi_{hop}(k_x, k_y)$, the parameter $\varepsilon_F$ establishes the onside interaction inside the same atomic orbital. If $\xi_{hop}(k_x, k_y)$ term is added, then the coefficient "$t$" says about transition metals behavior that is more complicated. If both terms are included in order to perform the Fermi average of the ZTSC, the numerical procedure implicitly includes both, the narrow d-bands and wide s-bands behavior, such as the one found in layered HTSC cuprates. Therefore, this kind of numerical treatment should be able to infer, which properties strong depend on $\varepsilon_F$, when elastic scattering is included.

On the other hand, in nonequilibrium statistical mechanics, the ground state and the phase space for N particles are related quantities i.e., $E - E_0 = f\,N\,(<\varepsilon> - \varepsilon_0)$, since it contains the degrees of freedom $f$ and the averaged energy $<\varepsilon>$ linked to a quantum number for each particle. The configuration space, and the phase space are related spaces, since $<\varepsilon>$ includes an average using a distribution function which depends on time $f(t)$, and we can make used of the Boltzmann kinetic equation for the $\tau$-approximation (we called this a Wigner probabilistic distribution approach in our previous works, since it gives a classical window to the quantum world as Wigner probabilistic distributions are able to accomplish [45].

We have found that a space that links the two mentioned above spaces is the reduced phase space (RPS), where energy is conserved for the scattering processes. The RPS is given by two coordinates, i.e., the axes representing the real frequencies, and imaginary part of the zero temperature elastic scattering cross-section ($\Re[\widetilde{\omega}(\omega + i\,0^+)], \Im[\widetilde{\omega}(\omega + i\,0^+)]$). The RPS has some important properties for the nonequilibrium statistical mechanics, when we make use of the parameters "$l$" and "$\tau$"; for a gas of dressed fermion quasiparticles, because is possible to move from a complete description of a non-equilibrium state, to an abbreviated description, using a single distribution function for one quasiparticle. Additionally, a numerical analysis in the reduced phase space, finds the position in which the degrees of freedom change, indicating a transition between different physical phases, such as, the normal to the superconducting phase transition, and it is able to remarkably distinguish different physical behaviors within these phases (if we consider crystal group theory, alongside with a proper Fermi surface average of the elastic scattering cross-section, within the unitary limit for a 2D layer of Strontium-doped Lanthanum Cuprate).

## 5. Recommendations

The use of the reduced phase space based on the zero elastic scattering cross-section is a well-established procedure to study unconventional superconductors, where strontium atoms are present, as in Strontium-doped Lanthanum Cuprate, and Strontium Ruthenate.

The direct analysis of the RPS using group theoretical considerations allowed us in the research [44], to predict the Miyake-Narikiyo tiny gap [43] in triplet superconductors using a TB approach. In addition, we were able to find the limit in which point nodes can be found for Strontium Ruthenate with nonmagnetic disorder in the work [55], and also, to infer how the superconducting gap varies for a reduce phase space with quasinodal behavior in the article [21]. We also linked the RPS to the phase and the configuration spaces. For the Strontium-doped Lanthanum Cuprate, we analyzed the RPS for the three scattering regimes, depending on the scattering lifetime, and the mean free path in the research [34].

However, we think that a world opens to study the inclusion of several quantum mechanical properties given by the number of states when considering quantities, such as, the density of states using the reduce phase space within the zero temperature scattering cross-section formalism.

## References

[1] Bednorz, J. & Müller. K. (1986). Possible high $T_c$ superconductivity in the BaLaCuO system. Z. Physik B - Condensed Matter 64, 189–193. doi: 10.1007/BF01303701.

[2] M. Kastner, R. Birgeneau, G. Shirane, and Y. Endoh. 1998. Magnetic, transport, and optical properties of mono layer copper oxides Rev. Mod. Phys. 70, 897. doi: 10.1103/RevModPhys.70.89.

[3] Wu, M., Ashburn, J., Torng, C., Hor, P., Meng R., Gao, L. Huang, Z., Wang, Y. & Chu, C. (1987). Superconductivity at 93 K in a new mixed-phase Y-Ba-Cu-O compound system at ambient pressure. Phys. Rev. Lett. 58 (9), 908 doi: 10.1103/PhysRevLett.58.908.

[4] Bardeen, J., Cooper, L. & Schrieffer, J. (1957). Microscopic Theory of Superconductivity, Phys. Rev. 106 (1), 162. doi: 10.1103/PhysRev.106.162.

[5] Sheadem. T. (1994). Introduction to High Tc Superconductivity. Plenum Press.

[6] Waldran, I. (1996). Structure of Cuprate Superconductors. Wiley.

[7] Cava. I. (2000). Oxide Superconductors. J. Am. Ceram. Soc., 83:5, doi: 10.1111/j.1151-2916.2000.tb01142.x.

[8] Xiao, G. Streitz, F., Gavrin, Du, A. & Chien C. (1987). Effect of transition-metal elements on the superconductivity of Y-Ba-Cu-O, Phys. Rev. B.85 (16), 8782. doi: 10.1103/PhysRevB.35.8782.

[9] Ambegaokar, V. & Griffin, A. (1965) Theory of the Thermal Conductivity of Superconducting Alloys with Paramagnetic Impurities, Phys. Rev. 137 (4A), A1151. doi: 10.1103/PhysRev.137.A1151.

[10] Momono, M. & Ido, M. (1996) Evidence for nodes in the superconducting gap of $La_{2-x}Sr_xCuO_4$. $T^2$ dependence of electronic specific heat and impurity effects, Physica C 264, 311, doi: 10.1016/0921-4534(96)00290-0.

[11] Sun, Y. & Maki, K. (1995). Transport Properties of D-Wave Superconductors with Impurities, EPL 32, 355.

[12] Larkin, A. (1965). Vector pairing in superconductors of small dimensions. JETP Letters. Vol. 2 (5), 105. ISSN: 0370-274X.

[13] Pethick, C. & Pines, D. (1986). Transport processes in heavy-fermion superconductors. Phys. Rev. Lett. 57 (1), 118, doi: 10.1103/PhysRevLett.57.118.

[14] Takeya, J., Ando, Y, Komiya, S., & Sun, XF. (2002). Low-temperature electronic heat transport in $La_{2-x}Sr_xCuO_4$. Single crystals: unusual low-energy physics in the normal and superconducting states. Phys. Rev. Lett. 88 (7): 077001. doi: 10.1103/phys.rev.lett.88.077001.

[15] Scalapino, D. (1995) The case for d pairing in the cuprate superconductors, Physics Reports. 250 (6), 329 doi: 10.1016/0370-1573(94)00086-I.

[16] Hussey, N. (2002). Low-energy quasiparticles in High-Tc cuprates, Adv. in Phys, 51:8, 1685. doi: 10.1080/000187302110164638.

[17] Yamase, H. Sakurai, Y. Fujita, M. et al. (2021) Fermi surface in La-based cuprate superconductors from Compton scattering imaging. Nat Commun 12, 2223. doi: 10.1038/s41467-021-22229-6.

[18] Photopoulos, R. and Frésard, R. (2019), Cuprate Superconductors: A 3D Tight-Binding Model for La-Based Cuprate Superconductors Ann. Phys. 531, 1970044. doi: 10.1002/andp.201970044.

[19] Walker. M. B. (2001). Fermi-liquid theory for anisotropic superconductors. Phys. Rev. B. 64 (13) 134515. doi: 10.1103/PhysRevB.64.134515.

[20] Pitaevskii, L: (2008). Superfluid Fermi liquid in a unitary regime, Phys. Usp. 51, 603 doi: 10.1070/PU2008v051n06ABEH006548.

[21] Contreras, P., Osorio, D. & Devi, A. (2022). The effect of nonmagnetic disorder in the superconducting energy gap of strontium ruthenate, Physica B: Condensed Matter. Vol. 646, 414330. doi: 10.1016/j.physb.2022.414330.

[22] Reif, F. (1965). Fundamentals of Statistical and Thermal Physics. McGraw-Hill.

[23] Pitaevskii, L., Lifshitz, EM & Sykes, J. (1981). Physical Kinetics, Vol. 10, Pergamon Press.

[24] Dorfman, J. Van Beijeren, H. & Kirkpatrick, T. (2021). Contemporary Kinetic Theory of Matter. Cambridge University Press. doi: 10.1017/9781139025942.

[25] Kaganov, M. & Lifshitz, I. (1989). Quasiparticles: Ideas and Principles of Quantum Solid State Physics. 2nd edition. Moscow "Nauka".

[26] Mineev, V. & Samokhin, K. (1999). Introduction to Unconventional Superconductivity. Gordon and Breach Science Publishers.

[27] Schachinger, E. & Carbotte, J. (2003). Residual absorption at zero temperature in d-wave superconductors. Phys. Rev. B 67, 134509. doi: 10.1103/PhysRevB.67.134509.

[28] Lifshitz, I., Gredeskul S. & and Pastur, L. (1988) Introduction to the theory of disordered systems. John Wiley and Sons.

[29] Edwards, S. (1958). A new method for the evaluation of electric conductivity in metals, Philosophical Magazine, 3 (33) 1020. doi: 10.1080/14786435808243244.

[30] Ziman, J. (1979). Models of Disorder: The Theoretical Physics of Homogeneously Disordered Systems, Cambridge University Press.

[31] Contreras, P., Walker, M. B. & Samokhin, K. (2004). Determining the superconducting gap structure in $Sr_2RuO_4$ from sound attenuation studies below $T_c$. Phys. Rev. B, 70: 184528. doi: 10.1103/PhysRevB.70.184528.

[32] Contreras, P. (2011). Electronic heat transport for a multiband superconducting gap in Sr2RuO4. Rev. Mex. Fis. 57 (5) 395.

[33] Contreras, P. et al., (2014). A numerical calculation of the electronic specific heat for the compound $Sr_2RuO_4$ below its superconducting transition temperature. Rev. Mex. Fis. 60 (3) 184.

[34] Contreras, P. & Osorio, D. (2021) Scattering Due to Non-magnetic Disorder in 2D Anisotropic d-wave High Tc Superconductors. Engineering Physics 5, 1, doi: 10.11648/j.ep.20210501.11.

[35] Contreras, P. & Moreno, J. (2019). A non-linear minimization calculation of the renormalized frequency in dirty d-wave superconductors. Canadian Journal of Pure and Applied Sciences. Vol. 13 (2), 4765 ISSN: 1920-3853.

[36] Tsuei, C. & Kirtley, J. (2000). Pairing symmetry in cuprate superconductors. Rev. Mod. Phys., 72: 969. doi: 10.1103/RevModPhys.72.969.

[37] Landau, L. & Lifshitz, E. (1980). Statistical Physics. Pergamon Press.

[38] Contreras, P. & Osorio, D. (2023). A Tale of the Scattering Lifetime and the Mean Free Path. arXiv: 2301.05322 [cond-mat.supr-con] doi: 10.485550/arXiv.230105322.

[39] Blatt, F. (1957). Theory of mobility of electrons in solids, Academic Press.

[40] Schrieffer, J. (1970). What is a quasiparticle? Journal of Research of the National Bureau of Standards, Vol. 74A (4), 537.

[41] Davydov, A. (1965). Quantum Mechanics. Pergamon Press.

[42] Kvashnikov, I. (2003). The theory of systems out of equilibrium, 3rd Vol. Moscow State University Press.

[43] Miyake, K. & Narikiyo, O. (1999). Model for Unconventional Superconductivity of $Sr_2RuO_4$. Effect of Impurity Scattering on Time-Reversal Breaking Triplet Pairing with a Tiny Gap. Phys. Rev. Lett. 83, 1423. doi: 10.1103/PhysRevLett.83.1423.

[44] Contreras, P., Osorio, D. & Ramazanov, S. (2022). Nonmagnetic tight- binding effects on the γ-sheet of $Sr_2RuO_2$. Rev. Mex. Fis 68 (2) 1, doi: 10.31349/RevMexFis.68.020502.

[45] Carruthers, P. & Zachariasen, F. (1983). Quantum collision theory with phase-space distributions. oxides Rev. Mod. Phys. 55 (1), 245 doi: 10.1103/RevModPhys.55.245.

[46] Brandt, N. & Chudinov, S. (1975). Electronic structure of metals, Mir Publishers.

[47] Contreras, P. Osorio, D. & Beliayev, E. (2022) Dressed behavior of the quasiparticles lifetime in the unitary limit of two unconventional superconductors. Low Temp. Phys. 48, 187, doi: 10.1063/10.0009535.

[48] Contreras, P. Osorio, D. & Beliayev, E. (2022) Tight-Binding Superconducting Phases in the Unconventional Compounds Strontium-Substituted Lanthanum Cuprate and Strontium Ruthenate. American Journal of Modern Physics. Vol. 11 (2) 32, doi: 10.11648/j.ajmp.20221102.13.

[49] Yoshida, T. et al. (2012). Pseudogap, Superconducting Gap, and Fermi Arc in High-Tc Cuprates Revealed by Angle-Resolved Photoemission Spectroscopy. Journal of the Physical Society of Japan, 81: 011006, doi: 10.1143/JPSJ.81.011006.

[50] Kaganov, M. & Contreras, P. (1994) Theory of the anomalous skin effect in metals with complicated Fermi surfaces. Journal of Experimental and Theoretical Physics, 79: 985, 1994. ISSN: 0080-4630.

[51] Kaganov, M., Lyubarskiy, G. & Mitina, A. (1997) The theory and history of the anomalous skin effect in normal metals, Physics Reports, Vol. 288 (1–6), 291, doi: 10.1016/S0370-1573(97)00029-X.

[52] Baker, G. (2022) Non-local electrical conductivity in PdCoO2 (Ph.D. Thesis). University of British Columbia. doi: 10.14288/1.0421263.

[53] G. Baker, et al., (2022) Non-local microwave electrodynamics in ultra-pure PdCoO2 arXiv preprint arXiv: 2204.14239 doi: 10.48550/arXiv.2204.14239.

[54] Torkhov, N. et al., (2022) Conversion of the anomalous skin effect to the normal one in thin-film metallic microwave systems. Phys. Scr. 97 095809 doi: 10.1088/1402-4896/ac837d.

[55] Contreras, P., Osorio, D. & Tsuchiya, S. (2022) Quasi-point versus point nodes in $Sr_2RuO_2$, the case of a flat tight binding γ sheet. Rev. Mex. Fis 68 (6), 060501 1–8. doi: 10.31349/RevMexFis.68.060501.